\documentclass[a4paper,11pt]{article}
\usepackage[utf8]{inputenc}
\usepackage{amsmath,amsthm,amsfonts}
\usepackage{graphicx}
\usepackage{natbib}
\usepackage{color}

\def\st{s_t}

 \def\sr{s_r}
\def\sij{s_{ij}}
\def\srj{s_{rj}}

\def\stj{s_{tj}}

\def\d{\mathbf{d}}

\def\dt{d_t}

\def\taujSq{\mathbf{\tau}_j^2}
\def\tauSq{\boldsymbol{\tau}^2}

\def\epsilonrj{\varepsilon_{rj}}

\def\epsilonij{\mathbf{\varepsilon}_{ij}}

\def\epsilontj{\varepsilon_{tj}}

\def\alphaj{\alpha_j}
\def\alphahatj{\hat{\alpha}_j}

\def\alphahatb{\hat{\boldsymbol{\alpha}}}
\def\alphab{\boldsymbol{\alpha}}

\def\betaj{\beta_j}
\def\betahatj{\hat{\beta}_j}
\def\betahatb{\hat{\boldsymbol{\beta}}}
\def\betab{\boldsymbol{\beta}}

\def\m{\mathbf{m}}
\def\mj{\mathbf{m}_j}
\def\mt{m_t}

\def\mtj{m_{tj}}

\def\mud{\boldsymbol{\mu}}

\def\muj{\mathbf{\mu}_j}

\def\a{\mathbf{a}}
\def\ai{a_i}
\def\at{a_t}
\def\ar{a_r}

\def\ig{\Gamma^{-1}}

\begin{document}

\section*{Title page}

{\bf PREPRINT; SUBMITTED FOR REVIEW, December 2012}

{\bf Title} Fully scalable online-preprocessing algorithm for short oligonucleotide microarray atlases

{\bf Running title} Online-RPA

{\bf Authors} Leo Lahti, Department of Veterinary Bioscience, University of Helsinki, Finland and Laboratory of Microbiology, Wageningen University, Netherlands; Aurora Torrente, European Bioinformatics Institute, Wellcome Trust Genome Campus, Hinxton, Cambridge, UK and Department of Material Science and Engineering, Universidad Carlos III de Madrid, Legan\'es, Spain; Laura L. Elo, Department of Mathematics and Statistics, University of Turku, Turku, Finland and Turku Centre for Biotechnology, Turku, Finland; Alvis Brazma, European Bioinformatics Institute, Wellcome Trust Genome Campus, Hinxton, Cambridge, UK; Johan Rung, European Bioinformatics Institute, Wellcome Trust Genome Campus, Hinxton, Cambridge, UK

{\bf Corresponding author} Leo Lahti, Department of Veterinary Bioscience, University of Helsinki, PO Box 66, FI-00014 University of Helsinki, Finland; Tel: +31 6 1673 7991; Email: leo.lahti@iki.fi; Fax: +31 317 483829

{\bf Keywords:} microarrays, online-learning, preprocessing, probabilistic analysis, robust probabilistic averaging

\newpage

\begin{abstract}


Accumulation of standardized data collections is opening up novel opportunities for holistic characterization of genome function. The limited scalability of current preprocessing techniques has, however, formed a bottleneck for full utilization of contemporary microarray collections. While short oligonucleotide arrays constitute a major source of genome-wide profiling data, scalable probe-level preprocessing algorithms have been available only for few measurement platforms based on pre-calculated model parameters from restricted reference training sets. To overcome these key limitations, we introduce a fully scalable online-learning algorithm that provides tools to process large microarray atlases including tens of thousands of arrays. Unlike the alternatives, the proposed algorithm scales up in linear time with respect to sample size and is readily applicable to all short oligonucleotide platforms. This is the only available preprocessing algorithm that can learn probe-level parameters based on sequential hyperparameter updates at small, consecutive batches of data, thus circumventing the extensive memory requirements of the standard approaches and opening up novel opportunities to take full advantage of contemporary microarray data collections. Moreover, using the most comprehensive data collections to estimate probe-level effects can assist in pinpointing individual probes affected by various biases and provide new tools to guide array design and quality control. The implementation is freely available in R/Bioconductor at http://www.bioconductor.org/packages/devel/bioc/html/RPA.html.

\end{abstract}

\newpage

\section{Introduction}

Accumulation of research data in in-house and public repositories, such as the ArrayExpress \citep{Parkinson2010} and Gene Expression Omnibus \citep{Barret05nar} has created data collections that include tens of thousands of microarray experiments originating from standardized measurement platforms \citep{Parkinson2010}, helping to overcome some of the inherent biases in meta-analyses involving multiple measurement platforms \citep{Kilpinen08, Rudy11}. By combining data from hundreds of individual studies and thousands of commensurable microarray experiments it is possible to obtain a more holistic picture of genome function and carry out detailed investigations of targeted model systems and diseases that can benefit from large sample sizes of the new data collections \citep{Lukk10}. A major portion of contemporary microarray collections originates from short oligonucleotide microarrays \citep{Lockhart96}, whose applications range from gene expression profiling \citep{Parkinson2010} to alternative splicing and microbial community analysis \citep{Brodie07, Rajilic-Stojanovic2009, Nikkila2010}. With nearly a million arrays in the ArrayExpress database, being able to combine and analyse very large sets of arrays together is a key challenge with a variety of applications \citep{Kohane12, Lukk10, Schmid12, Zheng-Bradley10}. 

Reliable preprocessing of the data is central for investigations. While multi-array preprocessing techniques that combine 
information across multiple arrays to quantify probe effects have led to improved preprocessing performance \citep{Bolstad03}, 
the applicability of the standard multi-array techniques, such as RMA \citep{Irizarry03rma}, GC-RMA \citep{Wu04}, MBEI \citep{Li01mbei} 
and PLIER \citep{Affy05}, has been limited to a few thousand arrays due to the considerable memory requirements associated with 
processing up to a million probe-level features per a single array. This has formed a bottleneck for comprehensive meta-analysis of 
contemporary microarray data collections. Scalable variants of the standard preprocessing algorithms have been recently developed to tackle these shortcomings. The scalable refRMA \citep{Katz06} and fRMA \citep{McCall2010} algorithms rely on pre-calculated probe effect 
terms that are estimated from restricted reference training sets and then extrapolated to preprocess further microarray data. The applicability of these methods has, however, been limited to few specific microarray platforms with pre-calculated probe effect terms; while the ArrayExpress database lists 203 distinct Affymetrix platforms, fRMA is currently available only for a single mouse platform and 2 human platforms \citep{McCall11}. No platform-independent preprocessing methods have been available to date to extract and utilize complete probe-level information in large-scale sample collections in a scalable manner.

We introduce a fully scalable probe-level model for multi-array preprocessing based on Bayesian online-learning to overcome the key limitations of the current approaches. We have extended the Robust Probabilistic Averaging framework introduced in \citep{Lahti11rpa} to estimate probe affinities and to incorporate prior information of the probes in the analysis. This provides the basis for scalable online-learning based on sequential hyperparameter updates, allowing rigorous preprocessing of large microarray atlases on an ordinary desktop computer in small consecutive batches with minimal memory requirements and in linear time with respect to sample size. The proposed Online-RPA algorithm (RPA) provides the means to extract probe-level information across arbitrarily large data collections and it is readily applicable to all short oligonucleotide microarray platforms. Moreover, the analysis of probe performance can now be based on the most comprehensive collections of short oligonucleotide array data, which provides useful information also for microarray design and quality control. To our knowledge this is the only probe-level preprocessing method which is both fully scalable and platform-independent, providing new tools to take full advantage of contemporary microarray data collections.

\section{Results}

\subsection{Scalable preprocessing with online-learning: an overview}

The standard steps in microarray preprocessing include background
correction, normalization, and probe summarization. Probe-level
procedures that combine information across multiple arrays have been
found to improve preprocessing performance \citep{Bolstad03} but their
applicability to large sample collections has been limited due to huge
memory requirements associated with increasing sample sizes. The
available solutions have been based on learning and extrapolation of
probe-level effects from smaller reference training sets
\citep{Katz06, McCall2010}. We propose an alternative online-learning
procedure that can extract probe-level information across the complete
microarray collection with minimal memory requirements and in linear
time with respect to sample size based on Bayesian hyperparameter
updates. Assuming that appropriate microarray quality controls have
been applied prior to the analysis, the standard steps of background
correction, normalization, and probe summarization are applied to
consecutive batches of the data in four sweeps over the complete data
collection:

{\it Step 1: Background correction} In the first step, each individual array is background-corrected 
with the standard RMA background correction \citep{Irizarry03rma}. The processed data can be stored temporarily
on hard disk to speed up preprocessing in the following steps that operate on background-corrected data.

{\it Step 2: Quantile basis estimation} The base distribution for quantile normalization is obtained as the average over sorted probe-level signals from background-corrected data \citep{Bolstad03}. We have implemented a scalable version of standard quantile normalization where the initial base distribution is calculated from the first batch and updated at each new batch 
by taking weighted average between the current base distribution and the one from the new batch, weighted by their corresponding sample sizes. The final base distribution is identical to the one which would be obtained by jointly normalizing all arrays in a single batch. Similar approach is used in parallel implementations of RMA where the quantile basis from individual batches are calculated and applied to the complete data collection after averaging over the individual basis distributions \citep{Schmidberger09}.

{\it Step 3: Hyperparameter estimation} The key novelty of our approach is in introducing the online-learning model for consecutive probe-level hyperparameter updates. The background-corrected batches are normalised with the quantile base distribution, log-transformed, and finally used to update the model parameters. At the first batch, the model can be initialized by giving equal priors for the probes if no probe-specific prior information is available. The probe-level hyperparameters are then updated at each new batch, providing priors for the next batch. The final probe-level parameters are obtained after scanning over all batches in the data collection. Ideally, this is expected to yield an identical result with a single batch approach.

{\it Step 4: Probe summarization} In the last step, the final probe-level parameters obtained from the third step are applied to summarize the probes in each batch, yielding the final preprocessed data matrix.

\subsubsection{The probe-level model}


Let us summarize the probe-level model for a fixed probeset with $J$ probes measured at $T+1$ arrays. The model is based on 
background-corrected, normalized, and log-transformed probe-level data. We assume a Gaussian model for probe effects, where the 
signal \(\sij\) of probe \(j \in \{1,\dots, J\}\) in sample \(i \in \{1,\dots, T+1\}\) is modeled as a sum of the underlying target signal \(a_i\) and Gaussian mean and variance parameters \(\muj\), \(\taujSq\) that are directly interpretable as constant affinity \(\muj\) and stochastic noise \(\epsilonij \sim N(0,\taujSq)\), respectively: 

\begin{equation}
  \sij = \ai + \muj + \epsilonij.
\label{eq:model}
\end{equation}

In the following, let us describe the estimation procedure for the model parameters \(\a = [a_1, \dots, a_{T+1}]\), \(\mud = [\mu_1, \dots, \mu_J]\), \(\tauSq = [\tau_1^2, \dots, \tau_J^2]\), and the additional modeling assumptions that are necessary to circumvent unidentifiability of the probe affinity terms.

\subsubsection{Incorporating prior information of the probes}

We start the analysis by estimating the variance parameters \(\tauSq\), following the procedure in \citep{Lahti11rpa}. In summary, the analysis is based on probe-level differential expression signal between each sample \(t = [1, \dots, T]\) and a randomly selected reference sample \(r\). Then, given Eq.~\ref{eq:model}, the unidentifiable affinity parameters \(\muj\) cancel out, yielding \(\stj - \srj = (\at - \ar) + (\epsilontj - \epsilonrj)\). Denoting \(\mt = \st - \sr\), \(\dt = \at - \ar\), and applying the vector notation \(\m = [m_1, \dots, m_T]\), \(\d = [d_1, \dots, d_T]\), the full posterior density for the model parameters \(\d, \tauSq\) is obtained with the Bayes' rule as

\begin{equation}
  P(\d, \tauSq | \m) \sim P(\m | \d, \tauSq) P(\d,\tauSq). 
\label{eq:fullpost}
\end{equation}

Assuming independent observations \(\mj\), given the model parameters, and marginalizing over \(\epsilonrj\), the likelihood term in Eq.~\ref{eq:fullpost} is then given \citep{Lahti11rpa} by 

\begin{equation}
\begin{split}
P(\m | \d, \tauSq) = \prod_{tj} \int N(\mtj | \dt - \epsilonrj, \taujSq) N(\epsilonrj | 0, \taujSq) d\epsilonrj \\
\sim \prod_j (2 \pi \taujSq)^{-\frac{T}{2}} exp(- \frac{\sum_t (\mtj -\dt)^2 - \frac{ [\sum_t (\mtj - \dt)]^2} {T+1}}{2 \taujSq}).
\end{split}
\label{eq:datalikelihood}
\end{equation}


With non-informative priors for \(P(\d, \tauSq)\) the posterior of Eq.~\ref{eq:fullpost} would reduce to maximum-likelihood-estimation of Eq.~\ref{eq:datalikelihood} as in \citep{Lahti11rpa}. The Bayesian online-learning version, developed and validated in this paper, takes full advantage of the prior term. This forms the basis for sequential updates of the posterior in Eq.~\ref{eq:fullpost}. Assuming independent prior terms, a non-informative prior \(P(\d) \sim 1\), and inverse Gamma conjugate priors for the variances with probe-specific hyperparameters \(\alphaj\) and \(\betaj\) \citep{Gelman03}, the prior takes the form

\begin{equation}
  P(\d, \tauSq) =  P(\d)P(\tauSq) \sim \prod_j \ig(\taujSq; \alphaj, \betaj).
\label{eq:prior}
\end{equation}

The posterior in Eq.~\ref{eq:fullpost} is now fully specified given the likelihood (Eq.~\ref{eq:datalikelihood}), the prior (Eq.~\ref{eq:prior}), and the hyperparameters \(\alphab = [\alpha_1, \dots, \alpha_J]\), \(\betab = [\beta_1, \dots, \beta_J]\). 

Our primary interest is in estimating the probe-specific variances \(\tauSq\), while \(\d\) is a nuisance parameter that could be marginalized out from the model to obtain more robust estimates of \(\tauSq\). Since no analytical solution is available and sampling-based marginalization approaches would slow down computation, we find a single point estimate for the joint posterior in Eq.~\ref{eq:fullpost} as a fast approximation. Point estimates are found by iterative optimization of \(\d\) and \(\tauSq\). A mode for \(\d\), given \(\tauSq\), is searched for by a standard quasi-Newton optimization method \citep{Goldfarb70}. Then, given \(\d\), the priors (\(\alphab, \betab\)) and sample size \(T + 1\), the variance \(\taujSq\) follows inverse Gamma distribution with hyperparameters \(\alphahatj = \alphaj + \frac{T}{2}\) and \(\betahatj = \betaj + \frac{1}{2} (\sum_t (\mtj - \dt)^2 - \frac{(\sum_t (\mtj - \dt))^2}{T+1}))\), yielding

\begin{equation}
  P(\taujSq | \m, \d) \sim \Gamma^{-1}(\taujSq | \alphahatj, \betahatj).
\label{eq:tauprior}
\end{equation}

The point estimate for \(\taujSq\) in this model is readily given by the mode at \(\taujSq = \betahatj/(\alphahatj + 1)\). We give equal weight for all probes by initializing the process with \(\taujSq = \beta/(\alpha+1)\) with \(\alpha = \beta = 1\) for all \(j\) if specific prior information is not available. The parameters \(\d\) and \(\tauSq\) are then iteratively updated until convergence (\(<0.01\) change in parameter values in our experiments). The inverse Gamma hyperparameters corresponding to the final \(\tauSq\) can then be retrieved as \(\alphahatj = \alphaj + \frac{T}{2}\) and \(\betahatj = \taujSq (\alphahatj + 1)\).

\subsubsection{Online-learning of variance hyperparameters}

The above formulation allows incorporation of prior information of the probes in the analysis and sequential updates where the updated hyperparameters \(\alphahatb\), \(\betahatb\) from the previous batch provide priors for the next batch through Eq.~\ref{eq:fullpost} and the prior in Eq.~\ref{eq:tauprior}. In the absence of prior information we shall give equal weight for all probes \(j\) at the first batch by setting \(\alphaj = 1\); \(\betaj = 1\) for all \(j\). The hyperparameters \(\alphahatb\), \(\betahatb\) are then updated with new observations at each batch until the complete data collection has been screened through, yielding the final probe-level hyperparameters.


\subsubsection{Affinity estimation and probe summarization}

The remaining task after learning the probe-specific variance hyperparameters is to estimate the probeset-level signal \(\a\) and probe affinities \(\mud\) in Eq.~\ref{eq:model}. Unidentifiability of probe affinities is a well-recognized issue in microarray preprocessing \citep{Irizarry03rma}, and further assumptions are necessary to formulate an identifiable model. A standard approach, used in the widely used RMA algorithm \citep{Irizarry03rma} is to assume that the probes capture the underlying signal correctly on average and the probe affinities sum to zero: \(\Sigma_j \muj = 0\). We propose a more flexible probabilistic approach where this hard constraint is replaced by soft priors that keep the expected probe affinities at zero and allow higher deviations for the more noisy probes with a higher variance \(\taujSq\). To implent this we apply a Gaussian prior \(\muj \sim N(0, \taujSq)\) for the affinities in Eq.~\ref{eq:model}. This allows higher fluctuations for the more noisy probes, yielding a better fit between the probeset-level signal estimate \(\a\) and the less noisy probes with smaller \(\taujSq\). The prior expectation of the sum of probe affinities, \(\Sigma_j \muj\), will remain at zero but in contrast to RMA, deviations from this are allowed if supported by the data. Alternatively, the affinity priors could be determined based on known probe-specific factors, such as GC-content which is a key element in probe affinity estimation in the GC-RMA algorithm \citep{Wu04}. As probe performance is affected by a number of factors, however, we prefer the data-driven approach which can accommodate various, potentially unknown probe-level noise sources. Point estimates for the probe affinities are identified based on the following procedure. From Eq.~\ref{eq:model} we have \(\ai = \sij - \muj - \epsilonij \sim N(\sij, 2\taujSq)\). A maximum-likelihood estimate for \(\ai\) is then obtained as a weighted sum of \(\sij\) over the probes \(j\), weighted by the inverse variances: \(\ai = \frac{1}{\Sigma_j \frac{1}{2\taujSq}} \Sigma_j \frac{1}{2\taujSq}(\sij)\). Given the estimated \(\ai\), the corresponding maximum-likelihood estimate for each \(\muj\) is then given by \(\muj^{(i)} = \sij - \ai\). Robust estimates of \(\muj\) are obtained by taking an average over affinity estimates across multiple samples, yielding the final estimates \(\mud = [\mu_1, \dots, \mu_J]\). This completes the specification of the model and allows probe summarization according to Eq.~\ref{eq:model}. Given \(\mud\), \(\tauSq\), the final estimate for the probeset-level signal \(\ai\) is now readily obtained by Eq.~\ref{eq:model} as the weighted sum of \(\sij - \muj\) over the probes \(j\), weighted by the inverse variances: \(\ai = \frac{1}{\Sigma_j \frac{1}{\taujSq}} \Sigma_j \frac{1}{\taujSq}(\sij - \muj)\).

\subsection{Validation}

In the following, let us investigate the scalability and parameter convergence of the new online-learning algorithm. The model performance is further validated by comparisons to alternative preprocessing methods with standard benchmarking procedures based on the AffyComp spike-in data sets \citep{Cope04} and a large-scale human gene expression atlas \citep{Lukk10}.

\subsubsection{Scalability and parameter convergence}


Ideally, the online-learning procedure is expected to yield identical results with the standard single-batch algorithm. We confirmed this by comparing the results obtained with the regular single-batch and online-learning versions of the RPA algorithm for a moderately sized data set of 300 randomly selected samples, which can be preprocessed with both models. The probeset-level signal estimates (\(\a\)) correlated to a high degree (Pearson correlation \(r > 0.995; p < 10^{-6}\)) between the single-batch and online-learning versions. The selected batch size may affect the running time depeding on the available memory resources but the results were robust to varying batch sizes of 10, 50, 100 and 300 samples (data not shown); we have used a batch size of 50 samples in the experiments. The high correspondence between the single-batch and online-learning models confirms the technical validity of the online version. 

Parameter convergence depends on the versatility of the data collection and overall probe-level noise, which can vary between probesets \citep{McCall2010}. For certain probesets, the probe parameters start to convergence before 2000-3000 samples or later, indicating that the sample sizes of \(\sim 1000\) arrays typically applied with fRMA \citep{McCall2010, McCall11} may in some cases be be too low to ensure parameter convergence (Supplementary Fig.~1). 

The scalability of Online-RPA was investigated by preprocessing up to 20,000 HG-U133A CEL files from ArrayExpress. The model scales up linearly with respect to sample size (Supplementary Fig.~2), with 10 hours preprocessing time for 20,000 CEL files on a standard desktop (Z400 with four 3.06 GHz processor cores). 

\subsubsection{Comparison with alternatives}

The currently available approaches for scalable preprocessing can be classified in (i) traditional single-array preprocessing methods and (ii) frozen multi-array techniques. In single-array algorithms, the preprocessing is performed separately for each array. Such approaches are fully scalable but cannot combine probe-level information across multiple arrays, limiting their accuracy compared to multi-array procedures \citep{Bolstad03}. We include the MAS5 algorithm \citep{Affy05} as the reference as this is one of the most well-known single-array preprocessing techniques. The frozen multi-array techniques include the refRMA \citep{Katz06} and fRMA \citep{McCall2010} algorithms. To our knowledge, fRMA \citep{McCall2010}, which incorporates ideas from refRMA \citep{Katz06}, is the only available multi-array preprocessing algorithm for scalable preprocessing of large-scale microarray collections. The fRMA is based on a standardized database of pre-calculated probe effect terms, which are then applied to preprocess new arrays in the database. The estimation procedure for probe effects is not scalable, however, and applicability of fRMA is currently limited to three Affymetrix platforms. In addition to the standard ``single-array'' fRMA model (fRMA), we consider a second variant which includes an additional model for batch effects (fRMA-batch). This incorporates additional experiment-specific information to the analysis which cannot be utilized by the other methods. While this can improve preprocessing performance, batch information is not necessarily available for heterogeneous microarray collections, and sets further requirements for the application of the fRMA-batch procedure. Finally, we include in the comparisons the widely used RMA algorithm \citep{Irizarry03rma}. In contrast to the other approaches RMA is not fully scalable, but the RMA preprocessed version of the \cite{Lukk10} data set is readily available, and as one of the most widely used standard preprocessing algorithms this is a relevant reference.

\subsubsection{Spike-in experiments}

\begin{figure}[h!]
\begin{center}
\begin{tabular}{ll}
\rotatebox{0}{\includegraphics[width=0.8\textwidth]{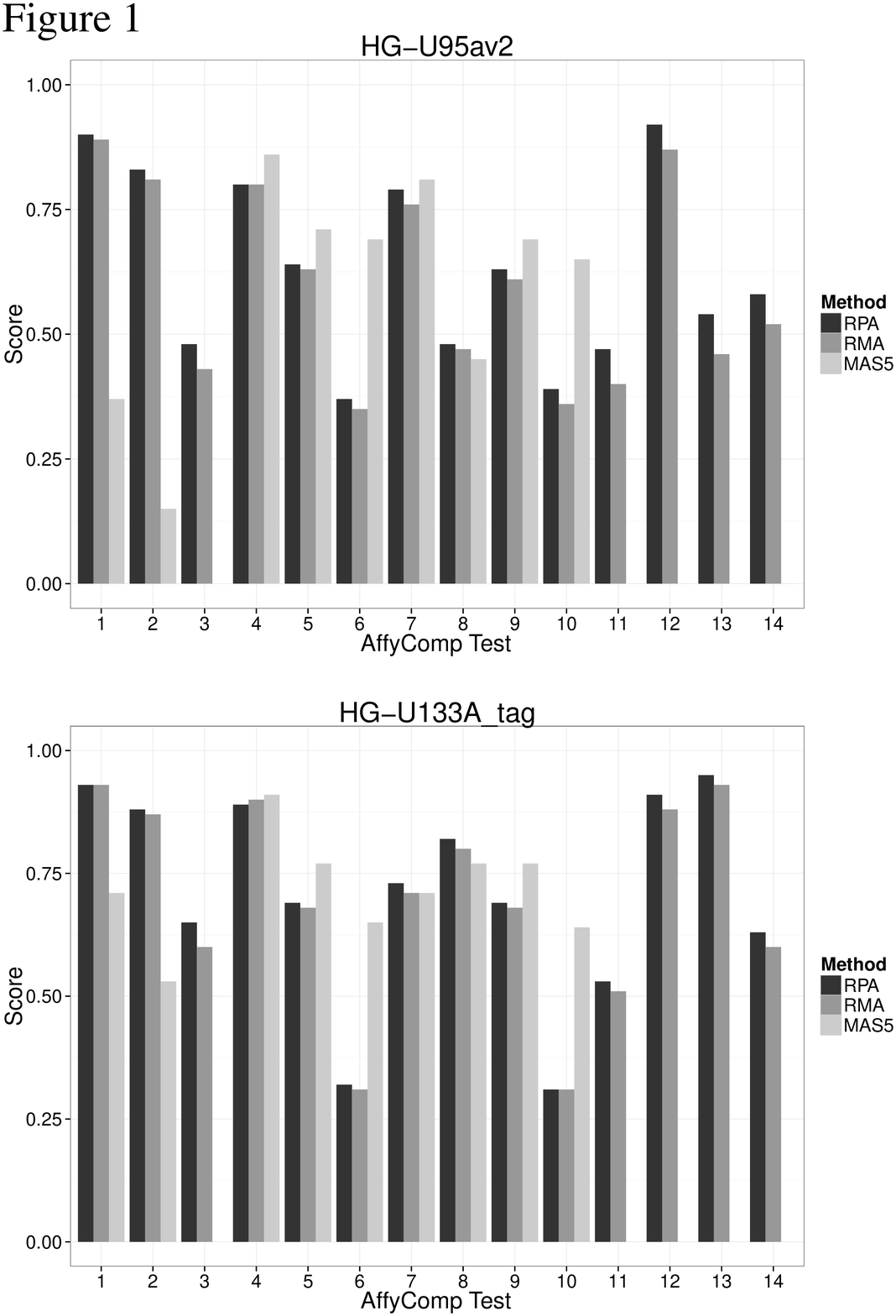}}
\end{tabular}
\end{center}
\caption{}
\label{fig:affycomp}
\end{figure}

The AffyComp website (http://affycomp.biostat.jhsph.edu; \cite{Cope04}) provides standard benchmarking tests for microarray preprocessing based on spike-in experiments on the Affymetrix HG-U95av2 and HG-U133A\_tag platforms. The tests are designed to quantify the relative sensitivity and accuracy of the alternative preprocessing algorithms based on known target transcript concentrations. We focus here on the scalable MAS5, fRMA, and RPA algorithms since most other methods at the AffyComp website have been designed for moderately sized data sets and have therefore a different scope than the scalable approaches. However, fRMA is not applicable to the spike-in data sets as pre-calculated probe-effect vectors are not available for these microarray platforms. Therefore we have included in the comparisons the widely used RMA algorithm which has a closely related probe-level model and is the key preprocessing method in the field. All arrays were preprocessed in a single batch with RPA. The Figure~\ref{fig:affycomp} summarizes the 14 AffyComp benchmarking tests for MAS5, RMA, and RPA. The complete comparison results for various preprocessing algorithms are available the AffyComp website\footnote{http://affycomp.biostat.jhsph.edu/AFFY3/comp\_form.html}. 

\begin{figure}[h!]
\begin{center}
\begin{tabular}{cc}
\rotatebox{0}{\includegraphics[width=0.9\textwidth]{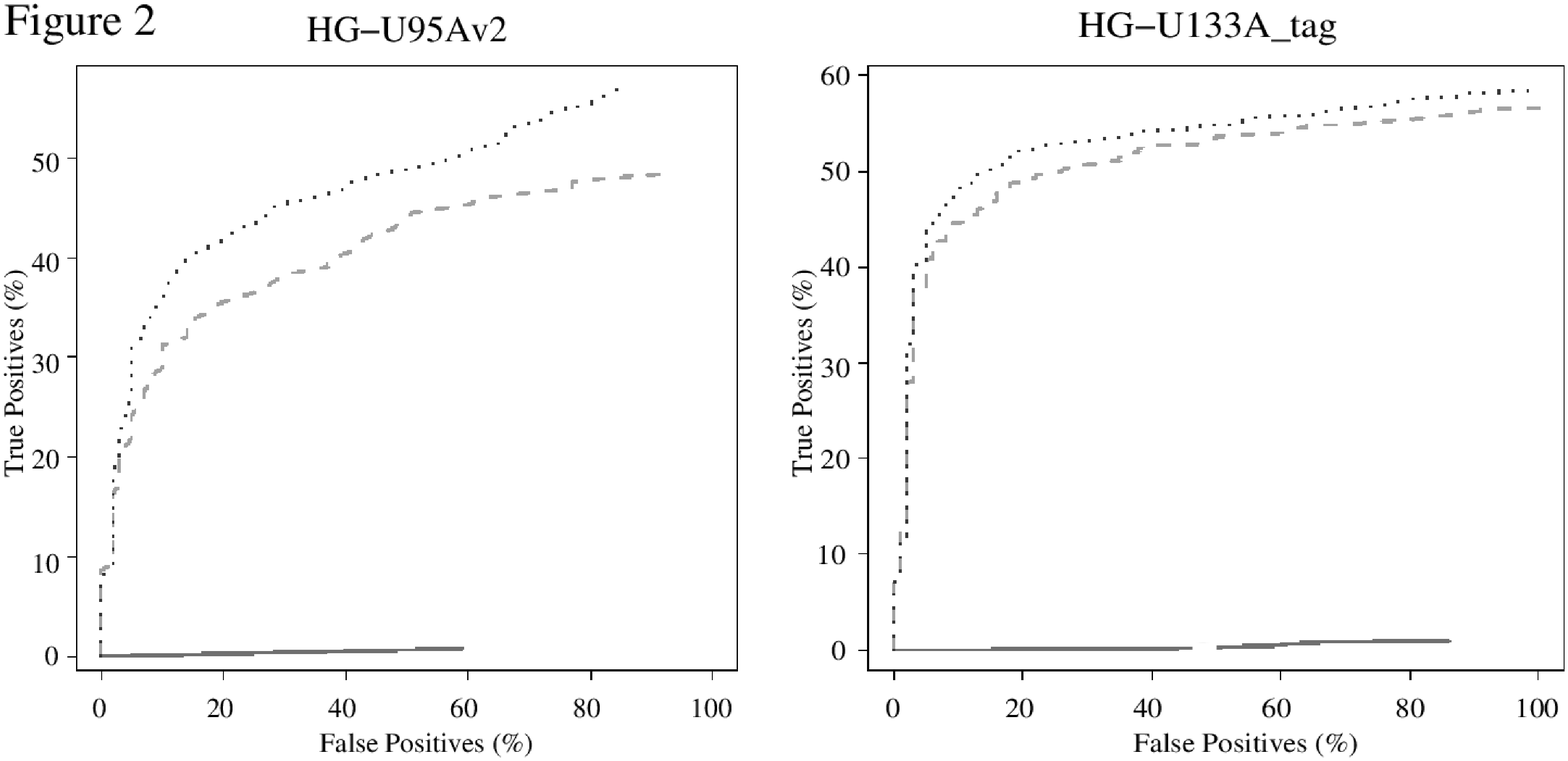}}
\end{tabular}
\end{center}
\caption{}
\label{fig:spikeroc2}
\end{figure}

RPA outperformed RMA in 13 and 11 tests (out of 14) on the HG-U95Av2 and HG-U133A\_tag data sets, respectively (Figure~\ref{fig:affycomp}). In particular, RPA outperformed RMA with respect to bias (AffyComp tests 5-10; Supplementary Fig.~3) and true positive-false positive rate, quantified by ROC/AUC analysis (AffyComp tests 11-14); the differences between RPA and the other methods were particularly salient with low concentration targets (Fig.~\ref{fig:spikeroc2}). In the other benchmarking tests, RPA and RMA had comparable performance. Interestingly, MAS5 had the smallest bias although RPA and RMA in general outperformed this method, and in certain tests such as ROC/AUC analysis MAS5 failed to distinguish the spike-in transcripts from noise. Certain methods in general outperform RPA at the AffyComp benchmarking tests, including the FARMS \citep{Hochreiter06} and GCRMA \citep{Wu03} algorithms. These methods have a more limited scalability than RPA, however, and hence a different scope. 

In summary, the fully scalable RPA had a similar or improved preprocessing performance in spike-in data sets compared to the standard RMA and MAS5 algorithms, and a wider scope than the only scalable probe-level alternative, fRMA, which is not available for the spike-in platforms \citep{McCall11}.

\subsubsection{Classification performance}

\begin{figure}[h!]
\begin{center}
\rotatebox{0}{\includegraphics[width=0.9\textwidth]{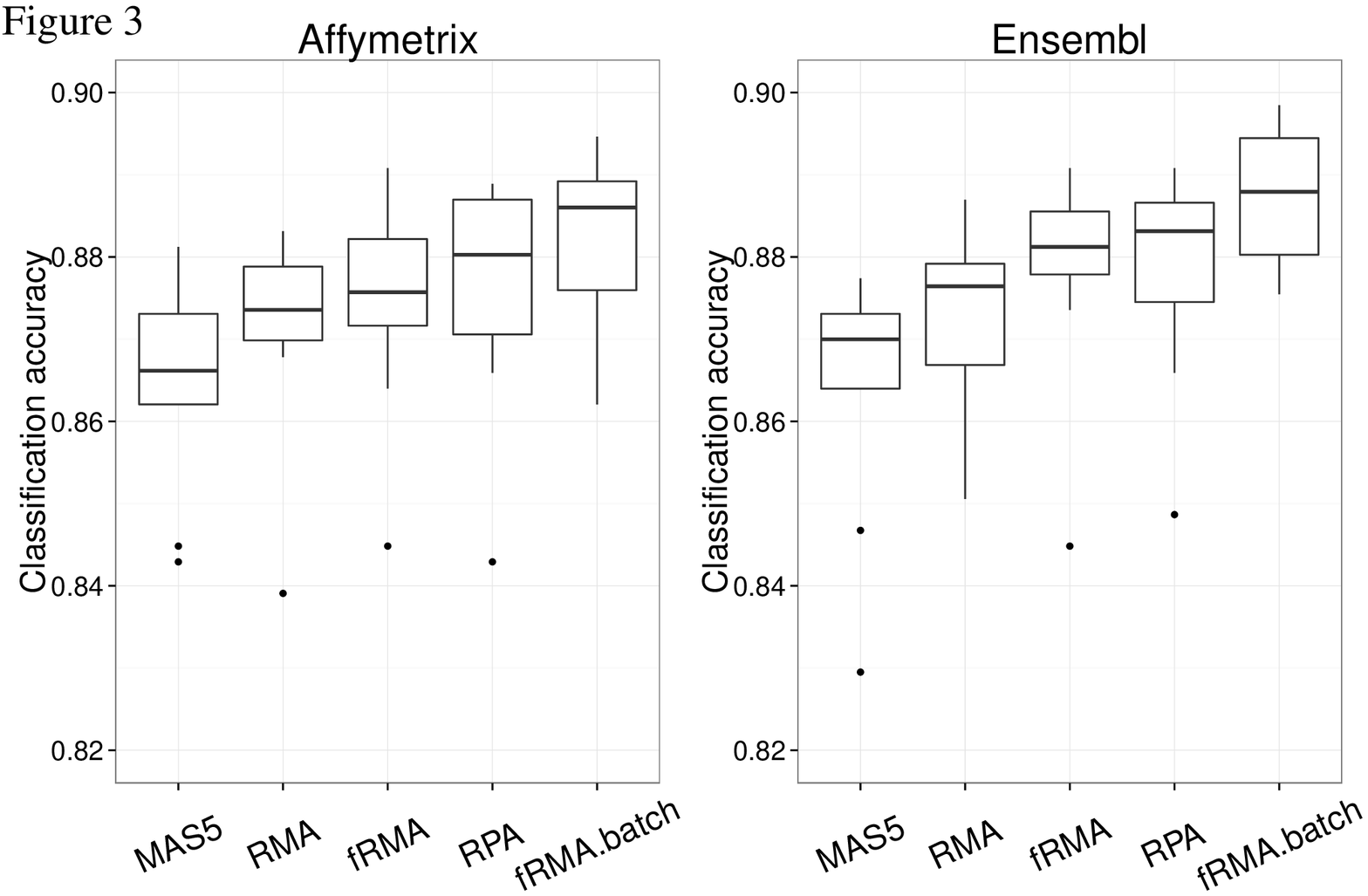}}
\end{center}
\caption{}
\label{fig:classification}
\end{figure}

We investigated the classification performance of the comparison methods based on the \citep{Lukk10} data collection of 5,372 samples with 369 cell and tissue types, disease states and cell lines (Fig.~\ref{fig:classification}). A random forest classifier \citep{Breiman01} was trained to distinguish between these classes based on 1000 randomly selected probe sets and 500 trees at 10 cross-validation folds, where the data was split into training (90\%) and test (10\%) sets. The singleton classes (150 samples) were excluded. The comparisons were performed with both the standard Affymetrix probe sets and alternative probesets based on the Ensembl Gene (ENSG) identifiers. RPA outperformed RMA and MAS5 (\(p<0.05\); paired Wilcoxon test). Differences between RPA and fRMA were not significant, and the fRMA with batch effect model (fRMA-batch) outperformed the other methods (\(p<0.05\)). It is important to note, however, that fRMA and fRMA-batch algorithms have a considerably more limited scope than Online-RPA, as detailed in Discussion. Further details on the data, class and batch definitions are provided in Materials and Methods.

\subsubsection{Correlation between technical gene replicates}

The standard Affymetrix arrays contain multiple probesets for certain transcripts. As the final benchmarking test, we compared the probeset-level summaries for such technical gene replicates on the \citep{Lukk10} data set, following \cite{Elo05, Lahti11rpa}. Pearson correlation was calculated for each Affymetrix probeset pair sharing the same EnsemblID (Bioconductor package {\it hgu133a.db}). The average correlations over all pairs were as follows: MAS5 0.46, RMA 0.53, fRMA 0.51, fRMA.batch 0.55 and RPA 0.54. The differences between the methods were significant (paired Wilcoxon test \(p < 0.01\)). In this comparison, RPA outperformed MAS5, RMA and fRMA (Supplementary Fig.~4).

\section{Discussion}

High memory requirements of the standard preprocessing techniques for short oligonucleotide arrays have limited their applicability 
to moderately sized data sets. Scalable probe-level preprocessing techniques, the refRMA \citep{Katz06} and fRMA \citep{McCall2010}, have been available only for few microarray platforms based on precalculated probe effects from restricted reference training sets. The lack of scalable, general-purpose preprocessing algorithms is hence forming a bottleneck for large-scale meta-analyses and full utilization of contemporary microarray collections. Our proposed method provides novel tools for taking advantage of the full information content in these data collections. With nearly a million arrays in the ArrayExpress database, being able to combine and analyse very large sets of arrays together is a key challenge with a variety of applications ranging from gene expression profiling \citep{Lukk10, Kohane12, Parkinson2010, Schmid12, Zheng-Bradley10} to alternative splicing and microbial community analysis \citep{Brodie07, Rajilic-Stojanovic2009, Nikkila2010}. 

We have introduced the first fully scalable online-learning algorithm that can extract and utilize individual probe effects across arbitrarily large  microarray collections, overcoming the key scalability limitations of the current preprocessing techniques. The algorithm learns probe-level parameters by sequential hyperparameter updates, processing the data in small consecutive batches. This makes the model readily applicable to contemporary microarray collections with tens of thousands of samples by extending the probabilistic framework introduced in \cite{Lahti11rpa}. In contrast to the alternatives our model is fully scalable and readily applicable to all microarray platforms as its application does not depend on precalculated probe effect terms. The model scales up in linear time with respect to sample size, with 10 hours running time for 20,000 CEL files in our experiments. The running time could be further accelerated by optimizing the implementation, using more efficient processors and parallelizing with multiple cores \citep{Schmidberger09}. Interestingly, we noticed that averaging of the probes, weighted by the probe-specific variance estimates provided by RPA, yielded similar results with the full model that additionally includes probe affinities. The differences between the two models were not significant, indicating that affinity analysis could be omitted to further speed up computation without compromising preprocessing performance. In analogy to fRMA, our model also allows the utilization of estimated model parameters as priors to preprocess further data sets. This can provide the advantages of single-chip methods and fRMA of not having to recompute the whole preprocessing procedure when new arrays are included in the data collection. Providing frozen parameter estimates can not only speed up computations but also allow reproducible analysis of single arrays for diagnostic and other purposes, as suggested in \citep{McCall2010}. 

Probe performance is affected by RNA degradation, non-specific hybridization, GC- and SNP-content, annotation errors and other, potentially unknown factors. While modeling of the probe effects have been shown to yield improved estimates of the target signal \citep{Irizarry03rma, Li01mbei}, the various sources of the probe-level noise remain poorly understood. While our model can be applied to learn the probe-level parameters based on subsets of the data, with the fully scalable extension the analysis of probe performance can now be based on the most comprehensive collections of short oligonucleotide array data that are available in the public repositories. As such, RPA can assist in nailing down individual probes affected by various biases, giving tools to guide microarray preprocessing and probe design in future studies and industry standards \citep{Katz06}. 

The widely used RMA algorithm is a special case of our single-batch model, assuming that the affinities sum to zero and all probes within a probeset have identical variances. The recent scalable extension, fRMA \citep{McCall2010}, has a more detailed model for probe-specific variances. While RPA was comparable to or outperformed the standard fRMA algorithm in our experiments, the fRMA-batch which utilizes additional sample metadata outperformed RPA. It is important to note, however, that the modeling of batch effects in fRMA is only possible when sufficient sample metadata is available which is not always the case with large and heterogeneous microarray collections, which are the primary target of scalable preprocessing algorithms. Moreover, batch effects could be modeled in a separate step for instance based on linear models \citep{Chen2011a, Leek2012}. Comparison of the various batch effect modeling techniques is out of the scope of this paper, however. Finally, the applicability of the fRMA and fRMA-batch algorithms is currently limited to only three Affymetrix platforms: HG-U133Plus2.0, HG-U133A, and MG-430 2.0 \citep{McCall11}, and could therefore not be included in the standard AffyComp spike-in benchmarking comparisons which rely on other microarray platforms. Therefore fRMA and fRMA-batch have a more limited scope than RPA. While certain methods, such as FARMS \citep{Hochreiter06} and GCRMA \citep{Wu03} with more detailed probe-level models outperform RPA in spike-in experiments, their scalability and hence the scope are more limited. 

The introduced online-learning approach therefore remains currently the only fully scalable preprocessing algorithm that is readily applicable on all short oligonucleotide platforms, outperforms the standard RMA algorithm, and can be used for scalable probe-level analysis and preprocessing of gene expression atlases involving tens of thousands of samples. This provides new tools to scale up investigations and to take advantage of the full information content of the rapidly expanding microarray data collections.

\section{Methods and Data Access}

\subsection{Data}

We have selected for comparisons a reasonably large, well-annotated and quality assessed microarray data set including 5372 human samples from a versatile collection of 369 cell and tissue types, disease states and cell lines from 206 public experiments and 162 laboratories, measured with the Affymetrix HG-U133A microarray \citep{Lukk10}. The biological groups are of varying sizes, and include 150 classes with only one sample (singleton classes); the annotations describing the group of each sample in the data set can be retrieved from the ArrayExpress archive (accession number: E-MTAB-62)\footnote{http://www.ebi.ac.uk/gxa/experimentDesign/E-MTAB-62}. This data set is ideal for benchmarking of scalable preprocessing methods since the alternative fRMA preprocessing model depends on the availability of precalculated probe effect terms, which are available for this microarray. Moreover, sufficient sample metadata is available for this data set to include batch effects in the fRMA model. In addition, despite of the heterogeneous origin of the data set, which made unfeasible to obtain `batches' in strictly the same manner as defined in \citep{McCall2010}, we could approximate them with the following approach. For each array within each experiment in the data set we retrieved the creation date of the CEL file from its HEADER section, under the DatHeader TAG, and assigned to the same batch those arrays from the same experiment (and laboratory) that were scanned on the same day. Thus, it was possible to assess the two available versions of fRMA, specifically the  ``single-array'' and ``batch-of-arrays''. Moreover, the sample size allows comparisons with the standard RMA preprocessing method \citep{Irizarry03rma}. Finally, in addition to the standard Affymetrix probe sets we have included in the comparisons alternative probe sets based on updated Ensembl gene mappings available through the hgu133ahsensgcdf (14.1.0) annotation package \citep{Dai05}. The reference probe effects for fRMA and the alternative mapping of probes to genes were built using the Bioconductor frmaTools \citep{McCall11}.

\subsection{AffyComp spike-in experiments}

The score components in Figure~\ref{fig:affycomp} are as follows: (1) median SD across replicates; (2) Inter-quartile range of the log-fold-changes from genes that should not change; (3) 99.9\% percentile of the log-fold-changes if from the genes that should not change; (4) \(R^2\) obtained from regressing expression values on nominal concentrations in the spike-in data; (5) slope obtained from regressing expression values on nominal concentrations in the spike-in data; (6) slope from regression of observed log concentration versus nominal log concentration for genes with low intensities; (7) same for genes with medium intensities; (8) same for genes with high intensities; (9) slope obtained from regressing observed log-fold-changes against nominal log-fold-changes; (10) slope obtained from regressing observed log-fold-changes against nominal log-fold-changes for genes with nominal concentrations less than or equal to 2; (11) area under the ROC curve (AUC; up to 100 false positives) for genes with low intensity standardized so that the optimum is 1; (12) AUC for genes with medium intensities; (13) AUC for genes with high intensities; (14) a weighted average of the ROC curves 11-13 with weights related to amount of data in each class. For full details, see \cite{Cope04}.

\section{Acknowledgements}

LL received funding from the Finnish Alfred Cordelin foundation and Academy of Finland (decision 256950). AT was supported by the Ram\'on Areces Foundation. LLE was supported by the Academy of Finland (decisions 127575, 218591). JR was supported through funds from The European Community's Seventh Framework Programme (FP7/2007-2013), ENGAGE Consortium, grant agreement HEALTH-F4-2007-201413.  All authors have reviewed the manuscript.

\section{Disclosure declaration}

The authors do not have competing interests.

\newpage

\section{Figure Legends}

Figure~\ref{fig:affycomp}: The benchmarking statistics for the AffyCompIII spike-in data for RPA, RMA, and MAS5 for the HG-U95Av2 (top) and HG-U133A\_tag (bottom) platforms. RPA and MAS5 represent fully scalable algorithms, and the standard RMA algorithm has been included as a benchmark since its fully scalable extension, fRMA, is not available for the spike-in platforms. For clarity of presentation, we have transformed the scores 1-3 with \(1-x\); so that the score value of 1 corresponds here to ideal performance at all 14 scores. For a full description of the 14 benchmarking components, see Materials and Methods.

Figure~\ref{fig:spikeroc2}: Average ROC curves for low-abundance targets with nominal concentrations at most 4 picoMolar and nominal fold changes at most 2 in the AffyCompIII spike-in data for MAS5 (solid line), RMA (dashed line), and RPA (dotted line) on the HG-U95Av2 (left) and HG-U133A\_tag (right) platforms. The Figure has been adapted from AffyCompIII test 5C. For details, see \cite{Cope04}. 

Figure~\ref{fig:classification}: Classification performance in Lukk et al. data set \citep{Lukk10} for the comparison algorithms. The 5372 samples were classified into 369 cell and tissue types, and after excluding the singleton classes the classification performance was quantified by random forest classifier based on 10-fold cross-validation as described in the Results. Online-RPA outperforms RMA and MAS5 (\(p<0.05\)). Differences between RPA-online and fRMA are not significant, and fRMA-batch outperforms the other methods (\(p<0.05\)).

\newpage

\section{Figures}

\newpage


\section{References}

\vspace{0mm}


\end{document}